\documentclass[12pt,preprint]{aastex}

\newcommand{\um}{$\mu$m~}
\newcommand{\ums}{$\mu$m}

\def\nh {${\rm N_\mathrm{H}}$}

\def\ergs{\ifmmode {\rm\,ergs\,s^{-1}}\else
    ${\rm\,ergs\,s^{-1}}$\fi}
\def\ergscm2Hz{\ifmmode {\rm\,ergs\,cm^{-2}\,s^{-1}\,Hz^{-1}}\else
    ${\rm\,ergs\,cm^{-2}\,s^{-1}\,Hz^{-1}}$\fi}
\def\ergcm2s{\ifmmode {\rm\,ergs\,cm^{-2}\,s^{-1}}\else
    ${\rm\,ergs\,cm^{-2}\,s^{-1}}$\fi} 
\def\kmsMpc{\ifmmode {\rm\,km\,s^{-1}\,Mpc^{-1}}\else
    ${\rm\,km\,s^{-1}\,Mpc^{-1}}$\fi}
                       
\shorttitle{Spitzer Spectra for SWIRE Sources}
\shortauthors{Weedman et al.}

\begin{document}

\title{AGN and Starburst Classification from Spitzer Mid-Infrared Spectra for High Redshift SWIRE Sources}

\author{D. Weedman\altaffilmark{1}, M. Polletta\altaffilmark{2}, C. J. Lonsdale\altaffilmark{2}, B. J. Wilkes\altaffilmark{3}, B. Siana\altaffilmark{4}, J. R. Houck\altaffilmark{1}, J. Surace\altaffilmark{4}, D. Shupe\altaffilmark{4}, D. Farrah\altaffilmark{1}, H.E. Smith \altaffilmark{2}}

\altaffiltext{1}{Astronomy Department, Cornell University, Ithaca, NY 14853; dweedman@astro.cornell.edu}
\altaffiltext{2}{Center for Astrophysics and Space Sciences, University of California, San Diego, La Jolla, CA 92093}
\altaffiltext{3}{Harvard Smithsonian Center for Astrophysics, 60 Garden St., Cambridge, MA 02138}
\altaffiltext{4}{Spitzer Science Center, California Institute of Technology, 220-6, Pasadena, CA 91125}

\begin{abstract}
  
Spectra have been obtained with the Infrared Spectrograph (IRS) on the Spitzer Space Telescope for 20 sources in the Lockman Hole field of the SWIRE survey. The sample is divided between sources with indicators of an obscured AGN, based primarily on X-ray detections of optically-faint sources, and sources with indicators of a starburst, based on optical and near-infrared spectral energy distributions (SEDs) which show a luminosity peak from stellar photospheric emission.  Ten of the 11 AGN sources have IRS spectra which show silicate absorption or are power laws; only one AGN source shows PAH emission features.  All 9 of the sources showing starburst SEDs in the near-infrared show PAH emission features in the IRS spectra.  Redshifts are determined from the IRS spectra for all 9 starbursts (1.0 $<$ z $<$ 1.9) and 8 of the 11 AGN (0.6 $<$ z $<$ 2.5).  Classification as AGN because of an X-ray detection, the classification as AGN or starburst derived from the photometric SED, and the IRS spectroscopic classification as AGN (silicate absorption) or starburst (PAH emission) are all consistent in 18 of 20 sources.  The surface density for starbursts which are most luminous in the mid-infrared is less than that for the most luminous AGN within the redshift interval 1.7 $\la$ z $\la$ 1.9. This result implies that mid-infrared source counts at high redshift are dominated by AGN for f$_{\nu}$ (24$\mu$m) $\ga$ 1.0\,mJy.

\end{abstract}


\section{Introduction}

Imaging surveys with the Spitzer Space Telescope ($Spitzer$) using the Multiband Imaging Photometer (MIPS) \citep{rie04} detect sufficient sources at 24 $\mu$m to resolve the cosmic background \citep{hau98} at mid-infrared wavelengths \citep{pap04, dol06}.  The source counts have been modeled to explain the infrared background as arising from the evolution of luminous, star-forming galaxies (starbursts) whose luminosity is primarily in the infrared and which arises from emission by dust \citep{lag04, chr04}. It is also known from various infrared and X-ray studies that there are many obscured Active Galactic Nuclei (AGN) \citep{ale01}, so many of the Spitzer mid-infrared sources could be dusty AGN instead of starbursts. It is essential to understand how the mid-infrared sources divide between starbursts and AGN so that the cosmic star formation history and the evolution of AGN can be distinguished and described as a function of redshift. This is a necessary step in understanding the relationships among the formation and evolution of stars, galaxies, and the massive black holes powering AGN within dusty environments. 

Extensive spectroscopic studies with the Infrared Space Observatory (ISO) indicated that the strength of emission features from polycyclic aromatic hydrocarbons (PAH emission) is the primary mid-infrared diagnostic of starbursts \citep{gen98, lut98, lau00}.  Relative to the underlying continuum, the PAH features from starbursts are at least 10 times stronger than from AGN, and this diagnostic leads to the conclusion that most ultraluminous infrared galaxies (ULIRGS) are powered by starbursts \citep{rig00}.  If the $Spitzer$ surveys at 24 $\mu$m are dominated by faint and distant ULIRGS which are starbursts, this conclusion implies that strong PAH features should frequently be seen in these sources.  So far, this expectation has not been confirmed with $Spitzer$ spectroscopic observations. 
  
To begin understanding the faint 24 $\mu$m source population, spectra have previously been obtained with the Infrared Spectrograph on $Spitzer$ (IRS) \citep{hou04} of faint sources discovered with MIPS surveys.  Some of these sources are invisible optically, to limits fainter than $\sim$ 26 mag in $Bw$, $R$, and $I$, so this population of highly obscured sources had not been discovered before $Spitzer$. Initial results of this spectroscopy are in \citet{hou05}, \citet{yan05}, \citet{lut05}, \citet{wee06a}, and \citet{wee06b}. Redshifts were determined from mid-infrared spectral features for 43 of 58 sources observed in the Bo\"{o}tes field \citep{hou05,wee06b}, for 6 of 8 sources selected by infrared color criteria in the $Spitzer$ First Look Survey (FLS) field \citep{yan05}, for 14 of 18 radio sources in the FLS \citep{wee06a}, and for 2 of 2 sources selected as submillimeter sources by \citet{lut05}.  In summary, of the 86 mid-infrared spectra of faint $Spitzer$ 24\,\um sources discussed to date (f$_{\nu}$(24\,\um) $\sim$ 1 mJy), redshifts were measured in the IRS spectra for 65 sources. The remaining 21 sources have featureless continua, to within the signal-to-noise ratio (S/N) of the IRS observations. 

The most important results so far are that the sources are generally at high redshift and have infrared spectra dominated by a strong silicate absorption feature at rest frame 9.7\,\um rather than by PAH emission.  For the 65 sources with IRS spectroscopic redshifts, the redshift (z) ranges from 1.1 to 2.8, with a median z = 2.1.  59 of these 65 sources have strong silicate absorption, with only 6 having redshifts derived from PAH emission.  These results are dominated by sources from the Bo\"{o}tes field \citep{hou05,wee06b}, which were all chosen only on the basis of extreme infrared to optical flux ratios.  (Defining IR/opt = $\nu$f$_{\nu}$(24\,\um)/$\nu$f$_{\nu}$($R$), the lower limit for this parameter in the Bo\"{o}tes  sources is 40.)  However, the optically-faint sources selected because they obey the radio-infrared correlation expected for starbursts also show IRS spectra characterised by silicate absorption \citep{wee06a}.  A few sources selected using the color criterion f$_{\nu}$(24\,\um)/f$_{\nu}$(8\,\um) $>$ 10 show PAH emission at z $\sim$ 2 \citep{yan05}, but only sources selected using submillimeter detections \citep{lut05} consistently show PAH emission characteristic of starbursts.  From existing observations, therefore, the optically faint, infrared-selected population appears to be primarily an AGN population, although this conclusion depends on the interpretation that sources with silicate absorption are obscured AGN.  Where are the high-redshift, luminous starbursts?  Why are they not common in samples observed so far?

The present study is designed to address these issues arising from the previous results. First, we attempt to validate the spectral classification that assigns AGN as the primary source of mid-infrared luminosity in sources observed to contain strong silicate absorption.  This is done by selecting a sample of sources independently classified as AGN, either because of X-ray luminosity or because their spectral energy distributions (SEDs) are power laws with no evidence of a stellar component.  Second, we attempt to verify that high redshift starbursts do indeed exhibit strong PAH emission by selecting a sample of starbursts independently of their mid-infrared spectra.  This is done by selecting sources which show a peak in the SED arising from stellar photospheric luminosity, as observed with the $Spitzer$ Infrared Array Camera (IRAC) \citep{faz04}.  Mid-infrared spectra of these samples obtained with the $Spitzer$ IRS are presented, and the overall spectral energy distributions (SEDs) are also discussed.  Finally, the space densities of starbursts found in this way are compared to the space densities of AGN at similar redshifts.

\section{Source Selection}

\subsection{Selection of AGN}

The Lockman Hole field has the most extensive multi-wavelength coverage of any field in the $Spitzer$ Wide-Area Infrared Extragalactic Survey (SWIRE) \citep{lon03, lon04}, which includes a 70ks image of 0.6 deg$^{2}$ obtained with the Chandra X-Ray Observatory (CXO). Details of the optical, infrared, and X-ray observations in the SWIRE Lockman field are given in \citet{pol06}, hereinafter P06.  

The best independent evidence that an optically-obscured $Spitzer$ source contains a luminous AGN arises if the source is identified as a luminous X-ray source (L(X) $>$ 10$^{42}$ ergs s$^{-1}$). All infrared-luminous AGN will not have observable X-ray counterparts even if they contain intrinsically luminous X-ray sources, because the X-rays can also be obscured in the most highly absorbed sources.  But selection of $Spitzer$ sources which also have the X-ray characteristics of AGN  provides a subset of infrared sources which can be confidentally classified as AGN-powered.  Therefore, we used the CXO survey to select in the Lockman field 8 X-ray sources which are also optically faint but sufficiently bright in the mid-infrared (f$_{\nu}$ (24\ums) $>$ 0.9\,mJy) for IRS spectroscopy.  These sources have broad-band (0.3-8 keV) CXO fluxes $\ga$ 10$^{-15}$ ergs cm$^{-2}$ s$^{-1}$, which lead to absorption-corrected X-ray luminosities L(X) $\ga$  10$^{43}$ ergs s$^{-1}$.  We applied the additional criterion that sources be optically faint ($R$ $\ga$ 22\,mag) so that the IR/opt would resemble the obscured sources in the previous IRS samples reviewed above.  This sample of 8 represents 50\% of the total number of sources in the CXO survey area satisfying these X-ray, infrared, and optical selection critieria. In addition, the brightest 24$\mu$m source which is an X-ray source was also observed (A2), although it is brighter than the optical limit for the other sources. Two objects already identified as high-redshift, Compton thick AGN in P06 are included in this sample (sources A1 and A4 in Table 1).  Characteristics of these SWIRE-CXO sources are in Tables 1 and 3 (sources A1 through A9).

By comparing X-ray sources with the multiwavelength SEDs determined with photometry from optical through infrared wavelengths, \citet{pol06} show that AGN can also be identified by the power-law shape of the SED from optical through infrared wavelengths.  Two sources in the Lockman field chosen as AGN in this way were also included for IRS observations in order to compare the AGN classification deriving from the SED with that from the mid-infrared spectrum.  These sources are also in Table 1 (sources A10 and A11).

\subsection{Selection of Starbursts}

Because of the availability of photometry from IRAC, the SED fits for SWIRE sources can locate objects with a well defined, redshifted peak in flux density at rest wavelength $\sim$1.6\,\um caused by the photospheric opacity minimum and resulting continuum maximum in giant or supergiant stars.  The clear presence of this peak is firm evidence that a stellar, photospheric component contributes to the near-infrared fluxes, and this peak can provide a photometric redshift estimate \citep{sim99}.  Using SEDs with a sufficiently dominant stellar component in near-infrared wavelengths to yield a photometric redshift, we have identified those sources within SWIRE survey fields which fall into various redshift ranges based on our estimates of the photometric redshift.  

For the present study, we selected for IRS observation a sample of sources in the full 9 deg$^{2}$ Lockman Field with f$_{\nu}$ (24\ums) $\ga$ 1.0\,mJy which were estimated to have a photometric redshift in the range 1.5 $<$ z $<$ 3.2.  Objects can be identified within this redshift range when the flux density in the IRAC 4.5\,\um band or 5.8\,\um band becomes brighter than in the other bands, because the rest-frame 1.6\,\um feature redshifts into the band.  We also applied an optical selection requiring $r$ $>$ 23 (although source B8, which was remeasured because it is close to a bright star, was subsequently found to be slightly brighter than this magnitude).   This redshift range was chosen in order to have a sample of faint $Spitzer$ 24\,\um sources chosen by stellar luminosity for comparison with the AGN samples previously reviewed \citep{hou05,wee06b} which are at z $\sim$ 2. Because of the luminous stellar component, we refer to these sources as starbursts during the remaining discussion.  The properties of the starburst sources are in Tables 2 and 4.  

Within the entire area of the Lockman Hole field with both infrared and optical coverage, there are 403 sources with f$_{\nu}$ (24\ums) $\ga$ 1.0\,mJy and $r$ $>$ 23.  Of these 403, 27 have the maximum IRAC flux within the 4.5\,\um band (expected redshift range 1.5 to 2.2) and 14 have the maximum IRAC flux within the 5.8\,\um band (expected redshift range 1.8 to 3.2), for a total of 41 sources which should be within 1.5 $<$ z $<$ 3.2, or $\sim$ 10\% of the entire flux-limited sample.  The sample in Tables 2 and 4 includes 9 sources (8 if B8 is excluded), or $\sim$ 20\% of the total number of sources within the redshift interval 1.5 $<$ z $<$ 3.2 that would satisy the selection criteria.

\section{Observations and Data Analysis}

IRS observations for the sources discussed in this paper are summarized in
Tables 1 and 2. 
Spectroscopic observations were made with the IRS Short Low module in
order 1 (SL1) and with the Long Low module in orders 1 and 2 (LL1 and
LL2), described in \citet{hou04}\footnote{The IRS was a collaborative venture between Cornell
University and Ball Aerospace Corporation funded by NASA through the
Jet Propulsion Laboratory and the Ames Research Center}.  These give low resolution spectral
coverage from $\sim$8\,$\mu$m to $\sim$35\,$\mu$m.  Sources were placed on
the slits by offsetting from a nearby 2MASS star.

All images when the source was in one of the two nod positions on each
slit were coadded to obtain the two-dimensional image of the source spectrum.  The background which was subtracted for LL1 or LL2 included coadded background images that added both nod positions having the source in the other slit (i.e., both nods on the LL1 slit when the source is in the LL2 slit produce LL1 spectra of background only) together with the alternative nod position in the same slit, yielding a background observation with three times the integration time as the integration time for the source.  This procedure reduces noise in the background but can only be used when LL1 and LL2 exposure times are the same.
The difference between coadded source images minus coadded background images was used for the spectral extraction, giving two independent extractions of the spectrum for each LL order.  These independent extractions were compared to reject any highly outlying pixels in either
spectrum, and a final mean spectrum was produced.  For SL1, there was no separate background observation with the source in the SL2 slit, so background subtraction
was done between coadded images of the two nod positions in SL1. The SWIRE-CXO sources were processed with version 11.0 of the SSC pipeline, because these observations were made earlier; other sources were processed with version 13.0. Extraction of
source spectra was done with the SMART analysis package \citep{hig04}.  

Use of the standard pipeline flux calibration requires extracting spectra of sources within the same pixels as applied to the flux calibration stars in the pipeline processing.  Perpendicular to dispersion, the number of pixels used in the "extraction width" for the standard flux extraction scales with spatial resolution and is typically set with an average value of 8 pixels FWHM at the central wavelength of an order.  This extraction width is several times greater than the spatial resolution of the spectra in order to assure that all flux is included.  For faint sources, dominated by background signal, S/N can be improved by restricting the number of pixels used to define the source spectrum.  To accomplish this, we have applied a typical extraction width of only 4 pixels.  While S/N is improved, some of the source flux in the outlying pixels is lost in this way, and we apply a correction to change the fluxes obtained with the narrow extraction to the fluxes that would be measured with a standard extraction.  This flux correction is derived by extracting an unresolved source of high S/N with both techniques and is a correction of about 10\%, although the correction varies with order and with wavelength.

All IRS spectra discussed in this paper are shown in
Figures 1 - 4; displayed spectra have been boxcar-smoothed to the approximate resolution of the different IRS modules (0.2\,\um for SL1, 0.3\,\um for LL2, and 0.4\,\um for LL1).  Smoothing is done because a resolution element contains about two pixels on the detector array. Figures 1 and 2 contain the IRS spectra of objects selected as AGN, and Figures 3 and 4 contain the IRS spectra of objects selected as starbursts.  
Photometric data with IRAC (3.6--8.0$\mu$m) and MIPS (24$\mu$m) from the 
SWIRE Survey are also shown with the spectra.  It can be seen from comparison of the spectral flux densities in Figures 1 - 4 with the MIPS 24\,$\mu$m fluxes shown in the Figures that the extracted spectra typically agree at 24\,\um to within 10 \% of the MIPS flux.   The complete SEDs of these objects are shown in Figures 5 and 6 for AGN and Figures 7 and 8 for starbursts.

\section{Discussion}

\subsection{Classification of Infrared Spectra}

Our objective is to use IRS spectra to determine the dominant mechanism producing the mid-infrared luminosity for a source, so it can be determined whether the source belongs to the AGN luminosity function or to the starburst luminosity function.  Certainly, it is overly simplistic to conclude that any given source has its infrared luminosity completely powered by either AGN or starburst, as demonstrated by numerous sources which are clearly composite \citep{far03}.  An excellent example is Markarian 231, which we use as a template for the mid-infrared spectrum 
of an absorbed AGN. It is the only local example (z $<$ 0.05) of a Compton-thick quasar, but its optical spectrum is a type 1 AGN while also showing extensive evidence of optical absorption \citep{gal02, gal05}.  The intrinsic X-ray luminosity is similar to the sources in Table 3 \citep{brt04}.  Yet, Markarian 231 has also long been known to show evidence of an extensive starburst, both in the optical \citep{bok77} and in the submillimeter and radio \citep{dow98, car98}. An extended molecular disk surrounding the AGN contains an obscured starburst which dominates the overall bolometric luminosity, because the bolometric luminosity arises primarily in the far infrared.  However, the near-infrared and mid-infrared luminosity arises from hotter dust that has been heated by the AGN \citep{kra97, mil96}. Markarian 231 is, therefore, the prototypical example of a source classifiable as AGN-powered at all wavelengths from X-ray through mid-infrared, but as starburst-powered at longer wavelengths.  Within mid-infrared source counts, an object with the characteristics of Markarian 231 should be assigned as an AGN, but in far-infrared source counts, assigned as a starburst.

Spectroscopic templates which define the spectral characteristics and which determine the redshift are illustrated in Figures 1 - 4 for each spectrum. These templates are from the IRS spectra shown in \citet{wee05}.  For simplicity, we illustrate only two templates for sources with spectral features: an AGN with strong silicate absorption defined by Markarian 231, and a starburst spectrum defined by NGC 3079, whose mid-infrared spectrum is dominated by PAH emission features.  Markarian 231 is used as the AGN template because it has known AGN characteristics and also has a silicate absorption feature with depth similar to that in the sources observed. (The Markarian 231 template shows a peak at rest-frame 8\,\um, labeled "Bump" in Figures 1 and 2, which is not emission but instead is the unabsorbed continuum having absorption features on either side of this peak, with silicate absorption to longer wavelengths and hydrocarbon and ice absorptions to shorter wavelengths \citep{spo04}.)  Sources A1 and A10 in Figures 1 and 2 show no spectral features, so the template illustrated for comparison is that of a featureless power-law derived empirically from a sample of QSOs observed by ISO and communicated to us by B. Schulz.  NGC 3079 is used as the starburst template because it is representative of a "pure" PAH spectrum, showing very little continuum at 10 $\mu$m, between the strongest PAH features. 

The results in the present paper apply only to the classification of sources based on their mid-infrared luminosity and mid-infrared spectral features, because our primary objective is to distinguish AGN and starbursts for sources found in surveys at 24 $\mu$m.
The clear result of the mid-infrared IRS spectra is that there is a straightforward spectroscopic distinction between objects classified as AGN because of their X-ray luminosities or power-law SEDs, and objects classified as starbursts because of the photospheric component in the SED.  Ten of the 11 AGN spectra shown in Figures 1 and 2 have either strong silicate absorption or a featureless power law, with only one source (A9) showing PAH features in addition to an absorbed continuum.  By contrast, all of the 9 starbursts in Figures 3 and 4 show distinctive PAH emission.  An optimal fit to source B6 would include both PAH emission and an absorbed component, so both the Markarian 231 template (silicate absorption) and NGC 3079 template (PAH emission) are shown for this source.   

The photometric SEDs of all sources are shown in Figures 5 - 8.  For comparison, template SEDs from known sources are also illustrated. These are not necessarily the optimal fits from a large set of templates, but are intended to illustrate differences between AGN and starburst templates.  The comparisons verify that the AGN or starburst classifications which would be derived from the IRS spectra are also consistent with those which would be derived from photometric SEDs.  

The SEDs of all AGN in Table 1 are in Figures 5 and 6.  The comparison templates are Markarian 231 \citep{berta05}, the Compton-thick quasar SWIRE\_J104409.95+585224.8 (Torus) from P06, a quasar template (TQSO1) from \citet{hat05} derived by combining quasars with large infrared/optical flux ratios in both the Sloan Digital Sky Survey and SWIRE, and a red quasar template (QSO2) from P06 derived from combining the
observed optical/near-IR spectrum of the red quasar FIRST J013435.7$-$093102 \citep{gregg02} with observed IR data from quasars with
consistent optical data. 

SEDs of all starbursts in Table 2 are in Figures 7 and 8. The signature of a starburst component is the redshifted "hump" in flux density at rest wavelength $\sim$1.6\,\um caused by the continuum maximum in giant or supergiant stars \citep{sim99}.  The template SED which is illustrated for a pure starburst is M82 \citep{silva98}.  

Some of the observed AGN and starburst SEDs in Figures 5 - 8 are also compared with composite templates which show both AGN and starburst components.  These composite templates are IRAS 22491$-$1808 (I22491) \citep{fritz05}, IRAS
22491$-$1808 (I20551) \citep{fritz05}, IRAS 05189$-$2524 (I05189) \citep{devriendt99}, and IRAS19254$-$7245 South
(I19254) \citep{berta03}.
 
Figures 5 and 6 show that 9 of the 11 AGN have SEDs that show only a power law, without the presence of a photospheric hump.  Only sources A7 and A9 show evidence of this stellar component in the IRAC bands.  All of the starburst sources in Figures 7 and 8 have SEDs with the near-infrared dominated by the photospheric component.  What is notable about the overall consistency in classification is that the AGN classification derived from a CXO detection, the classification derived from the photometric SED, and the IRS spectroscopic classification are all consistent in 18 of 20 sources as to whether the sources are dominated by AGN or starbursts.  Only one source classified as an AGN because of the X-ray luminosity shows measurable PAH emission, indicating that the mid-infrared spectrum contains a starburst component. This is source A9, which is illustrated in Figure 2 with both starburst and AGN templates for the IRS spectum.  Based on the SED fit, A9 would be classified as an AGN because the best fit
is obtained with a Seyfert 2 SED (Figure 6), although this SED includes a starburst component.  Of all 20 sources observed, therefore, this is the only one for which a classification derived strictly from the IRS mid-infrared spectrum (starburst because PAH features are present) would give a result disagreeing with the classification at another wavelength (AGN because of X-ray luminosity). (The starburst template is also shown in Figure 1 for source A6 as an illustration that possible emission features in this spectrum are not fit by the PAH template.)

As a further check on the distinctiveness of this classification, we derive an overall mean AGN spectrum and a mean starburst spectrum for these faint sources by combining all of the spectra of sources with confident redshifts and confident classifications to obtain average spectra.  The AGN average is determined from all sources in Table 1 except A9 (ambiguous classification) and A10 (no redshift); the starburst average is determined from all sources in Table 2.  The results are in Figures 9 and 10.  The most notable result of these combined spectra is the clear division between the absorption spectrum in the AGN, which fits the Markarian 231 prototype very well, and the PAH emission spectrum of the starbursts, which fits the NGC 3079 starburst prototype.  The mean AGN spectrum shows no indication of any PAH emission, and the mean starburst spectrum shows no indication of any absorption.

A spectroscopic classification derived from mid-infrared spectral features strictly applies only to the source of the mid-infrared dust luminosity.  As yet, we have few fluxes observed beyond 24\,\um for the sources in this paper, which means that the known SED rarely extends beyond $\sim$ 10\,\um in the rest frame of the source.  For local ULIRGS, the bulk of the luminosity is within far-infrared and submillimeter wavelengths, so models of the SEDs to determine the starburst and AGN components are weighted by the source of far-infrared and submillimeter luminosity \citep{far03}. How the mid-infrared luminosities relate to bolometric luminosities which include longer wavelengths is a general question needed to understand the total energy budget of the 24\,\um source population, but we do not address that question in the present paper.  

\subsection{Characteristics of AGN}

Our primary objective in observing the 9 sources in Table 1 with X-ray detections was to determine if the infrared spectra showed the silicate absorption which had previously been assumed to arise from AGN.  Figures 1 and 2 illustrates that silicate absorption is present in 6 of the 9 X-ray sources (A2, A4, A5, A6, A7, and A8) and probably present along with PAH emission in A9. The AGN in Figures 1 and 2 with redshifts are all fit by the Markarian 231 template, except for source A9 which appears composite and is shown with both AGN and starburst templates.  Redshifts for sources A5, A6, A7, A8, and A11  were determined from the Markarian 231 template fit. Sources A1, A2, A3 and A4 have templates shown at the optically-derived redshift, but the IRS spectrum provides confirmation for A2 and A4. Source A9 has a photometric SED with a starburst component, and a redshift is derived from the PAH features in Figure 2; this redshift agrees with the photometric redshift.  

An indication that there are distinctive characteristics of AGN among $Spitzer$ 24\,\um sources was found by comparing infrared colors for sources with X-ray detections to the colors of sources without X-ray detections. \citet{brn06} compare the IRAC and MIPS fluxes for sources in the Bo\"{o}tes survey with the $\sim$ 5\% of those sources also detected by a CXO survey of the Bo\"{o}tes field.  Brand et al. show that the AGN, classified by X-ray detection, are distributed around a value of log[$\nu$f$_{\nu}$ (24\,\um)/$\nu$f$_{\nu}$ (8\,\um)] = 0, whereas the remaining sources are distributed about a value of 0.5.  This result is used to imply that AGN-powered sources have flatter mid-infrared spectra, which would arise if dust temperatures are higher for AGN compared to starbursts.  This distinction in the overall infrared properties between samples known to contain AGN and samples without an AGN indicator is an important contribution to our confidence that the mid-infrared luminosity of sources containing an AGN arises as a consequence of the AGN. 

Tables 1 and 2 include values of the parameter log[$\nu$f$_{\nu}$ (24\,\um)/$\nu$f$_{\nu}$ (8\,\um)] for the AGN and starburst sources in our samples.  This ratio is illustrated for the AGN and starburst samples in Figure 11, and the values are consistent with the conclusion of Brand et al. that AGN spectra are flatter. The median ratio for the 9 Chandra-selected AGN is $\sim$ 0.3, with the greatest value at 0.68 (A7).  The median for the 9 starbursts, counting limits, is $\sim$ 1.0, with the smallest value being 0.56 (B8), barely overlapping the distribution for AGN. 

Because both the X-ray and infrared spectra show evidence of absorption, the absorbing columns can be compared for the X-ray absorbing clouds of ionized gas and the infrared absorbing clouds of dust. The X-ray spectra are used to derive \nh\ as in P06, and these are given in Table 3.  The depth of the silicate absorption can also be related to the column density \nh\ using the grain model of \citet{li01}.  This model predicts the extinction by dust as a function of wavelength, A($\lambda$), and relates that extinction to \nh.  Using the graphical relation in their Figure 16, \nh\ = 2 $\times$ 10$^{22}$A(9.7\,\um) cm$^{-2}$, for A(9.7\,\um) the extinction in magnitudes at the maximum depth of the silicate feature.  The extinction A(9.7\,\um) can be measured from the infrared spectra by locating the continuum level on either side of the silicate feature and connecting these continua to define a flux density at 9.7\,\um for an unabsorbed continuum, f(cont), which is then compared to the minimum flux density actually observed within the absorption feature, f(9.7). With these definitions, A(9.7\,\um) = 2.5 log[f(cont)/f(9.7)]. 

From Figures 1 and 2, sources A2, A3, A4, A5, A6, and A8 show silicate absorption features with depths similar to the 50\% absorption in the Markarian 231 template, and sources A7 and A11 show deeper absorption.  If 50\% of the continuum is extincted, A(9.7\,\um) = 0.75.  This means that a source like Markarian 231 requires \nh\ = 1.5 $\times$ 10$^{22}$ cm$^{-2}$ to explain the silicate absorption.  The X-ray absorption column density for the 6 sources in Table 3 which are similar to Markarian 231 in the silicate extinction ranges from 3 $\times$ 10$^{22}$ cm$^{-2}$ to 10$^{24}$ cm$^{-2}$.  In all cases, the \nh\ derived from the X-ray absorption exceeds that derived from the silicate absorption, by up to a factor of $\sim$ 100.  Smaller column densities measured in silicates compared to those measured in the X-ray are not unexpected, even if the absorption comes from the same cloud or clouds. The silicate absorption measures only the absorption between the cooler, outer portion of an emitting dust cloud and a warmer, inner portion where the continuum arises, whereas the X-ray absorption arises from the column through the entire cloud. In addition, the X-ray continuum arising from an AGN can also be absorbed  within clouds close to the AGN where the temperature is too high for dust to exist. 

The measurement of extinction in the silicate feature becomes insensitive to column density for \nh\ $\ga$ 10$^{23}$ cm$^{-2}$, so it is difficult to compare silicate and X-ray absorptions quantitatively for large values of extinction when using infrared spectra of poor S/N.  For example, source A7 shows silicate absorption that is the deepest of all sources in Figures 1 or 2, approaching a level of zero continuum at the depth of the feature, which would be  infinite extinction.  But the poor S/N makes it impossible to determine precisely how much extinction is present.  If 90\% of the continuum is extincted,  this would yield A(9.7\,\um) = 2.4, or \nh\ = 5 $\times$ 10$^{22}$ cm$^{-2}$.  This is much less than the \nh\ = 3 $\times$ 10$^{23}$ cm$^{-2}$ determined from the X-rays, but perhaps the silicate extinction has been underestimated. 

A specific comparison showing the large differences between absorption measured from X-rays and that measured from the silicate feature is for the two Compton-thick sources A1 and A4.  For these two sources, the X-ray absorption gives \nh\ $\sim$ 10$^{24}$ cm$^{-2}$.  Yet, source A1 shows no silicate absorption, and source A4 shows silicate absorption consistent with the 50\% depth in Markarian 231, or \nh\ $\sim$ 1.5 $\times$ 10$^{22}$ cm$^{-2}$.  For these two sources, therefore, the X-ray absorbing column density is $\ga$ 100 times greater than the silicate absorbing column density.  It can be concluded that the absorption column densities measured from silicate absorption are always less than those measured from X-ray absorption, but the quality of the infrared spectra and the limited size of our sample prevent a more quantitative conclusion.

\subsection{Characteristics of Starbursts}

The 9 starburst sources in Figures 3 and 4 all have redshifts that are derived from the fit to the PAH emission features in the IRS spectrum using the NGC 3079 template.  

For determination of the luminosity function of obscured starburst galaxies, luminosities of the PAH features are a crucial parameter because these features arise from a photodissociation region which scales with the ionizing luminosities of the stars.   The PAH features can be used, therefore, to determine the overall luminosities of starbursts and to relate these luminosities to characteristics at other wavelengths \citep{lut98, lau00, bra04, yan05, des06}.  Measuring the strengths of the weaker PAH features in spectra of poor S/N, as in Figures 3 and 4, is challenging.  The strongest PAH emission feature is at rest frame 7.7\,\um, but the underlying continuum that should be subtracted from the emission feature is difficult to define if there is no observed continuum baseline at longer wavelengths, which occurs if z $\ga$ 2.  A similar problem arises at even lower redshifts for the 11.3\,\um and 12.7\,\um features.  The PAH feature for which the underlying continuum is least affected by redshift would be the 6.2\,\um feature, but this feature is sufficiently weak that accurate measurement in individual sources is often difficult. 

For measuring the starburst luminosities of the sources in Figures 3 and 4, we measure the peak luminosity of the 7.7\,\um feature and attempt to subtract the level of the underlying continuum, similarly to the method used for ISO spectra \citep{rig00}.  Using the average starburst spectrum in Figure 10, because the continuum is difficult to measure in individual objects, the continuum beneath the 7.7\,\um feature as extrapolated between 5\,\um and 10\,\um is 0.6 mJy, and the average 7.7\,\um peak is 3.3 mJy.  Using the definition of line to continuum ratio (l/c) from \citet{rig00}, the feature itself (peak minus continuum) is 2.7 mJy, giving l/c = 4.5 for the average starburst spectrum.  Rigopoulou et al. quote 3.0 for the ISO average of starbursts and 2.0 for the ISO average of ULIRGS. This comparison indicates that the spectra in Figures 3 and 4 have stronger PAH strengths relative to the continuum (e.g. equivalent widths) than the objects classified as starbursts in \citet{rig00}, although such comparisons have large uncertainty because of uncertainty in defining the underlying continuum.  

Although equivalent widths of PAH features are not the same in all starbursts, a measurement of the flux at the peak of the 7.7\,\um feature is an approximate way to compare luminosities of starbursts and is particularly useful if the S/N is so poor that it is difficult to separate the emission feature from the underlying continuum, as in the individual spectra in Figures 3 and 4.  The luminosities of these spectra at the 7.7\,\um peak, expressed in $\nu$L$_{\nu}$ units and determined from the flux peaks in Figures 3 and 4, are listed in Table 4. The mean luminosity is 5.2 $\times$ 10$^{45}$erg s$^{-1}$. For comparison, the luminosities of the 7.7\,\um peak in the most luminous starburst galaxies found by \citet{yan05} are 7 $\times$ 10$^{45}$erg s$^{-1}$ (IRS9, z = 1.83) and 1.3 $\times$ 10$^{46}$erg s$^{-1}$ (IRS6, z = 2.4), and that found by \citet{des06} is 9 $\times$ 10$^{45}$erg s$^{-1}$ for a source at z = 1.34.  For the prototype starburst nucleus of NGC 7714, the peak luminosity is 4.4 $\times$ 10$^{43}$erg s$^{-1}$ \citep{bra04} and is 1.5 $\times$ 10$^{43}$erg s$^{-1}$ for the NGC 3079 template in this paper. These starburst sources within the Lockman field are, therefore, among the most luminous yet detected at mid-infrared wavelengths and are $\sim$ 100 times more luminous than local examples like NGC 7714 and NGC 3079, based on scaling to the peak luminosity at the 7.7\,\um PAH feature.    

The average starburst spectrum in Figure 10 emphasises a strong selection effect when seeking luminous starbursts using f$_{\nu}$ (24$\mu$m) as a criterion.  Because of the rest-frame peak of the spectrum at the 7.7\,\um feature, there is a selection in favor of sources at z $\sim$ 2, which places this peak flux within the 24\,\um MIPS filter bandwidth. The combination of uncertainties introduced by this selection effect and introduced by application of the photometric redshift estimator used to select sources mean that the space density of luminous starbursts derived from our results is uncertain.  The results in Table 2 mean that 8 starbursts have been discovered with f$_{\nu}$ (24$\mu$m) $>$ 1.0\,mJy in the range 1.7 $\la$ z $\la$ 1.9 within the 9 deg$^{2}$ of the Lockman Hole survey field, for a surface density $\sim$ 1 deg$^{-2}$.  This is a lower limit to the total number of starbursts with this flux density and redshift criterion, because the selection based on photometric redshifts was incomplete.  For example, the source selection described in section 2 indicates that we included only 20\% of all starbursts within 1.5 $<$ z $<$ 3.2 based on the IRAC source selection.  However, the result that most of the sources observed fell into the narrow redshift range 1.7 $\la$ z $\la$ 1.9 indicates that the selection based on 24\,\um flux favors this narrow redshift interval.  At present, the best limits we can determine, therefore, are that the surface density of starbursts with f$_{\nu}$ (24$\mu$m) $>$ 1.0\,mJy and 1.7 $\la$ z $\la$ 1.9 is between 1 deg$^{-2}$ and 5 deg$^{-2}$, but the favorable 24\,\um selection for 1.7 $\la$ z $\la$ 1.9 implies that the lower bound is more nearly correct.  Despite this uncertainty, it is useful to compare the surface density of luminous starbursts to that of infrared-luminous AGN previously discovered in the same redshift interval.   

The surface density of optically-classified AGN to the same limit of f$_{\nu}$ (24\,\um) $>$ 1 mJy has been determined.  Using spectroscopic redshifts, \citet{bro05} derive the redshift distribution and luminosity function of optically-classified type 1 quasars discovered in the MIPS 24\,\um Bo\"{o}tes survey, to a limit of f$_{\nu}$ (24\,\um) $>$ 1 mJy and $R$ $<$ 21.7.  Brown et al. conclude that the shape of the rest-frame 8\,\um infrared luminosity function for these type 1 quasars is the same as derived from optical surveys, and that the redshift peak of quasar space density is at the same redshift.  They find 20 quasars in the $\sim$ 8 deg$^{2}$ Bo\"{o}tes field in the interval 1.7 $\la$ z $\la$ 1.9. This result accounts only for the optically bright, unobscured quasars within a $Spitzer$ 24\,\um sample brighter than 1 mJy.  The surface density of obscured AGN in Bo\"{o}tes with $R$ $>$ 22 exceeds that for the unobscured type 1 quasars within the same redshift interval \citep{wee06b}. These previous results indicate, therefore, that there are at least $\sim$ 5 AGN deg$^{2}$ with f$_{\nu}$ (24\,\um) $>$ 1 mJy and 1.7 $\la$ z $\la$ 1.9.  Depending on the completeness correction for our starburst sample, this comparison between starbursts and AGN indicates that the high-redshift luminosity function of mid-infrared sources with f$_{\nu}$ (24$\mu$m) $>$ 1.0\,mJy has at least as many AGN as starbursts and may be dominated by AGN.  

\section{Summary and Conclusions}

Mid-infrared spectra obtained with the IRS on $Spitzer$ are presented for a sample of 20 sources with f$_{\nu}$ (24$\mu$m) $\ge$ 1.0\,mJy derived from the SWIRE survey of the Lockman Hole field.  Sources were selected as AGN because of X-ray luminosities (9 sources) or power-law SEDs (2 sources).  Sources were selected as starbursts (9 sources) because of a strong photospheric component in the near-infrared SEDs derived from $Spitzer$ IRAC observations which allowed a photometric redshift estimate.  New redshifts were determined from IRS spectra for 14 sources, with redshifts measured from strong silicate absorption features in 5 AGN sources (1.14 $<$ z $<$ 2.25) and from strong PAH emission features in 9 starburst sources (0.98 $<$ z $<$ 1.91) and one AGN source (z = 1.04). Ten of the 11 AGN sources have IRS spectra which show silicate absorption or are power laws; only one AGN source shows PAH emission
features.  All of the 9 sources selected because of starburst classification in the near-infrared SEDs show PAH emission features in the mid-infrared IRS spectra.  

These results indicate that optically faint sources ($R$ $\ga$ 22\,mag) with f$_{\nu}$ (24$\mu$m) $\ge$ 1.0\,mJy which contain either a dominant AGN or a dominant starburst as determined from independent classifications at X-ray, optical, and near-infrared wavelengths have a predictable spectral classification in the mid-infrared: the AGN sources have IRS spectra which show silicate absorption or are power laws, and the starburst sources show PAH emission features. The results also show that photometric SEDs derived with IRAC and MIPS fluxes are an efficient method to locate starbursts at z $\sim$ 1.8. 

The starburst galaxies found at high redshift are $\sim$ 100 times more luminous in PAH emission than local prototypes.  The sample of starburst galaxies gives the result that for f$_{\nu}$ (24$\mu$m) $>$ 1.0\,mJy and 1.7 $\la$ z $\la$ 1.9, starburst galaxies have a surface density between $\sim$ 1 deg$^{-2}$ and 5 deg$^{-2}$, depending on the selection effects that should be applied. This is the redshift interval in which selection at 24\,\um optimises detection of starbursts based on strong PAH emission.  By comparing the starburst sample to previous samples of AGN detected by $Spitzer$ to comparable limits of f$_{\nu}$ (24$\mu$m) $\ga$ 1.0\,mJy, it is concluded that at z $\sim$ 1.8, the most luminous mid-infrared sources are dominated by AGN.  Surface densities of AGN with f$_{\nu}$ (24\,\um) $>$ 1 mJy and z $\sim$1.8 exceed the surface densities of starbursts with f$_{\nu}$ (24\,\um) $>$ 1 mJy, even at redshifts where starburst selection is favored because of strong PAH emission within the 24\,\um band.    

For AGN, absorption in both X-rays and in the mid-infrared silicate feature is compared.  Column densities \nh\ derived from X-ray absorption are always found to be greater than the \nh\ derived from silicate absorption, by factors that can exceed 100.  There are no correlations within the small sample of sources among X-ray absorbing properties or X-ray luminosities compared to infrared spectral properties, such as infrared spectral slope or extinction in the silicate absorption feature.  However, the AGN have flatter infrared spectra compared to the starbursts as measured by the ratio $\nu$f$_{\nu}$ (24\,\um)/$\nu$f$_{\nu}$ (8\,\um).

\acknowledgments
We thank D. Devost, G. Sloan, and P. Hall for help in improving our IRS spectral analysis and S. Higdon and J. Higdon for their effort in developing the SMART analysis package.  We thank B. Schulz for providing ISO spectra of quasars.  MP and BW acknowledge financial support for CXO observations through NASA grant G04-5158A.
This work is based in part on observations made with the
Spitzer Space Telescope, which is operated by the Jet Propulsion
Laboratory, California Institute of Technology, under NASA contract
1407. The SWIRE project is part of the $Spitzer$ Legacy Science Program. Support for this work by the IRS GTO team at Cornell University was provided by NASA through Contract
Number 1257184 issued by JPL/Caltech.

\clearpage

\begin{deluxetable}{lc cc cccccccc}
\tabletypesize{\footnotesize}
\rotate
\tablecaption{Optical \& Infrared Properties of the AGN Sample\label{tab1}}
\tablewidth{0pt}
\tablehead{
\colhead{ID}&
\colhead{Source Name\tablenotemark{a}} &
\colhead{$z$} &
\colhead{$r$} &
\colhead{F(3.6)} & 
\colhead{F(4.5)} &
\colhead{F(5.8)} &
\colhead{F(8)} &
\colhead{F(24)} &
\colhead{Class} &
\colhead{(24/8)\tablenotemark{d}}&
\colhead{time\tablenotemark{e}}\\
\colhead{}& 
\colhead{}& 
\colhead{}& 
\colhead{(Vega)} &
\colhead{($\mu$Jy)} & 
\colhead{($\mu$Jy)} &
\colhead{($\mu$Jy)} &
\colhead{($\mu$Jy)} &
\colhead{($\mu$Jy)} &
\colhead{(SED)}  &
\colhead{}&
\colhead{s}
}
\startdata
 A1  &  SJ104406.30+583954.1  &    2.430\tablenotemark{b}  &    23.4  &    45 &    65 &  129 &  242 &  1190  & AGN &   0.21$\pm$0.03 &1440,360\\
 A2  &  SJ104407.97+584437.0  &    0.555\tablenotemark{b}  &    20.1  &   644 &   848 & 1140 & 1590 &  7390  & AGN &   0.19$\pm$0.03 &720,240\\
 A3  &  SJ104351.87+584953.7  &    0.609\tablenotemark{b}  &    23.1  &    19 &    25 &   44 &  106 &   940  & AGN &   0.47$\pm$0.04 &1440,360\\
 A4  &  SJ104409.95+585224.8  &    2.540\tablenotemark{b}  &    23.5  &    64 &   151 &  407 & 1070 &  4200  & AGN &   0.12$\pm$0.03 &720,240 \\
 A5  &  SJ104453.07+585453.1  &    1.89\tablenotemark{c}   &    24.7  &    71 &   135 &  285 &  471 &  1210  & AGN &$-$0.07$\pm$0.03 &1440,360\\
 A6  &  SJ104613.48+585941.4  &    2.10\tablenotemark{c}   &    22.9  &    27 &    34 &   58 &  122 &  1140  & AGN &   0.49$\pm$0.04 &1440,360\\
 A7  &  SJ104354.82+585902.4  &    1.14\tablenotemark{c}   &    22.3  &    47 &    38 &   27 &   69 &   990  & SB  &   0.68$\pm$0.04 &1440,360\\
 A8  &  SJ104528.29+591326.7  &    2.31\tablenotemark{c}   &    23.8  &    32 &    45 &   86 &  200 &  2490  & AGN &   0.62$\pm$0.03 &720,240\\
 A9  &  SJ104706.95+592011.8  &    1.04\tablenotemark{c}   &    22.0  &   141 &   136 &  133 &  160 &  1150  & AGN &   0.38$\pm$0.03 &1440,360\\
 A10 & SJ103916.79+585657.9   &                  \nodata   &    22.9  &    10 &    24 &   69 &  274 &  4860  & AGN &   0.77$\pm$0.03 &480,240\\
 A11 & SJ104314.93+585606.3   &    2.25\tablenotemark{c}   & $>$24.3  &     9 &    22 &   62 &  115 &   950  & AGN &   0.44$\pm$0.04 &1440,600\\
\enddata

\tablecomments{The IRS data for sources A1-A9 were processed with version
11.0 of the SSC pipeline, and those for sources A10-11 were processed with
version 13.0.  All flux density and magnitude values are from the SWIRE survey.  Uncertainties in flux densities are typically $\pm$ 5\% ; uncertainties in $r$ magnitudes are typically $\pm$ 0.04 mag.}
\tablenotetext{a}{SJ stands for SWIRE\_J. SWIRE\_JHHMMSS.ss+DDMMSS.s is the
official IAU source name for sources discovered in the SWIRE fields.}
\tablenotetext{b}{Spectroscopic $z$ from optical spectrum (P06); sources A1, A2 and A4 are type 2 AGNs with narrow emission lines of high ionization in their optical spectra, source A2 shows both the Balmer series in absorption 
and high-ionization narrow emission lines, indicating a post-starburst galaxy.}
\tablenotetext{c}{Spectroscopic $z$ from IRS spectrum; redshift uncertainties are typically $\pm$ 0.2.}
\tablenotetext{d}{Measure of infrared spectral slope between 8\,\um and 24\,\um, given as ratio log[$\nu$f$_{\nu}$(24\,\um)/$\nu$f$_{\nu}$(8\,\um)].}
\tablenotetext{e}{First number is total integration time for each order of the Long Low
  spectrum; second number is total integration time in Short Low order 1.}
\end{deluxetable}

\begin{deluxetable}{lc cc cccccccc}
\tabletypesize{\footnotesize}
\rotate
\tablecaption{Optical \& Infrared Properties of the Starburst Sample\label{tab2}}
\tablewidth{0pt}
\tablehead{
\colhead{ID}&
\colhead{Source Name\tablenotemark{a}} &
\colhead{$z$\tablenotemark{b}} &
\colhead{$r$} &
\colhead{F(3.6)} & 
\colhead{F(4.5)} &
\colhead{F(5.8)} &
\colhead{F(8)} &
\colhead{F(24)} &
\colhead{Class}&
\colhead{(24/8)\tablenotemark{c}}&
\colhead{time\tablenotemark{d}}\\
\colhead{}& 
\colhead{}& 
\colhead{}& 
\colhead{(Vega)} &
\colhead{($\mu$Jy)} & 
\colhead{($\mu$Jy)} &
\colhead{($\mu$Jy)} &
\colhead{($\mu$Jy)} &
\colhead{($\mu$Jy)} &
\colhead{(SED)}  &
\colhead{}&
\colhead{s}
}
\startdata
 B1 &   SJ104217.17+575459.2  &   1.91  &     23.1 &  42  &  47 &    46 & $<$40 & 1004  & SB & $>$0.92       &1440,600\\
 B2 &   SJ104731.08+581016.1  &   1.81  &  $>$24.3 &  40  &  57 &    54 & $<$40 & 1050  & SB & $>$0.94       &1440,600\\
 B3 &   SJ105405.49+581400.1  &   1.82  &     24.7 &  38  &  48 &    63 &    42 & 1150  & SB & 0.96$\pm$0.05 &1440,600 \\
 B4 &   SJ103837.03+582214.8  &   1.68  &     25.6 &  68  &  93 &    97 &    72 & 1030  & SB & 0.68$\pm$0.03 &1440,600\\
 B5 &   SJ103744.46+582950.6  &   1.88  &     23.5 &  77  &  94 &    98 &    78 & 1440  & SB & 0.79$\pm$0.03 &1200,480\\
 B6 &   SJ103809.18+583226.2  &   0.98  &     25.7 &  26  &  33 &    40 & $<$40 & 1060  & SB & $>$0.95       &1440,600\\
 B7 &   SJ103856.98+585244.1  &   1.88  &     26.2 &  26  &  46 &    48 & $<$40 & 1060  & SB & $>$0.95       &1440,600\\
 B8 &   SJ104839.33+592149.0  &   1.89  &     22.6 &  75  &  95 &   120 &   125 & 1360  & SB & 0.56$\pm$0.02 &1200,480\\
 B9 &   SJ104620.38+593305.1  &   1.84  &  $>$24.3 &  19  &  26 & $<$43 & $<$40 & 1360  & SB & $>$1.05       &1200,480\\
\enddata
\tablecomments{The IRS data for all B sources were processed with
version 13.0. of the SSC pipeline. All flux density and magnitude values are from the SWIRE survey.  Uncertainties in flux densities are typically $\pm$ 5\% ; uncertainties in $r$ magnitudes are typically $\pm$ 0.04 mag.}
\tablenotetext{a}{SJ stands for SWIRE\_J. SWIRE\_JHHMMSS.ss+DDMMSS.s is the
official IAU source name for sources discovered in the SWIRE fields.}
\tablenotetext{b}{Spectroscopic $z$ from IRS spectrum; redshift uncertainties are typically $\pm$ 0.2.}
\tablenotetext{c}{Measure of infrared spectral slope between 8\,\um and 24\,\um, given as ratio log[$\nu$f$_{\nu}$(24\,\um)/$\nu$f$_{\nu}$(8\,\um)].}
\tablenotetext{d}{First number is total integration time for each order of the Long Low
  spectrum; second number is total integration time in Short Low order 1.}
\end{deluxetable}

\topmargin=3cm
\begin{deluxetable}{l r rr rrc}
\tabletypesize{\scriptsize}
\tablecaption{X-ray Properties of the SWIRE/CXO sample\label{tab3}}
\tablewidth{0pt}
\tablehead{
\colhead{Source ID} & 
\colhead{HR}     &
\colhead{C(0.3-8 keV)\tablenotemark{a}}& 
\colhead{F(0.3-8 keV)\tablenotemark{b}}&
\colhead{\nh$^{obs}$\tablenotemark{c}} & 
\colhead{\nh$^{rf}$\tablenotemark{c}}  & 
\colhead{Log(L(X))\tablenotemark{d}}   \\
\colhead{}  & 
\colhead{} & 
\colhead{} & 
\colhead{} &
\colhead{(cm$^{-2}$)} & 
\colhead{(cm$^{-2}$)} & 
\colhead{(erg s$^{-1}$)}
}
\startdata 
 A1  &     0.61$^{+0.21}_{-0.23}$   &     21.1$\pm$6.4    &     3.96$\pm$1.20  &  4.0$^{+2.8}_{-1.2}$ &    99$^{+70}_{-30}$  &   45.74$^{+0.22}_{-0.46}$  \\
 A2  &     0.10$^{+0.17}_{-0.16}$   &     48.0$\pm$8.4    &     9.29$\pm$1.63  &  1.7$^{+0.6}_{-0.5}$  &     5$^{+2}_{-2}$   &   43.44$^{+0.15}_{-0.24}$  \\
 A3  &     0.61$^{+0.17}_{-0.17}$   &     44.4$\pm$8.4    &     8.20$\pm$1.54  &  4.4$^{+2.2}_{-1.5}$  &    15$^{+8}_{-8}$   &   43.69$^{+0.19}_{-0.34}$ \\
 A4  &     0.85$^{+0.06}_{-0.39}$   &     11.3$\pm$4.9    &     1.87$\pm$0.81  &  7.4$^{+1.9}_{-4.9}$  &   200$^{+50}_{-130}$&   45.64$^{+0.25}_{-0.68}$  \\
 A5  &  $-$0.06$^{+0.07}_{-0.10}$   &    117.2$\pm$12.2   &    21.03$\pm$2.19  &  1.2$^{+0.3}_{-0.3}$ &    19$^{+5}_{-5}$   &   45.31$^{+0.11}_{-0.15}$  \\
 A6  &  $-$0.16$^{+0.07}_{-0.09}$  &    150.6$\pm$13.4   &    24.94$\pm$2.22  &  0.8$^{+0.2}_{-0.2}$  &    15$^{+4}_{-4}$   &   45.42$^{+0.12}_{-0.16}$  \\
 A7  &     1.00$^{+0.00}_{-0.46}$   &      9.0$\pm$4.4    &     1.57$\pm$0.78  &  3.5$^{+0.0}_{-1.7}$ &    26$^{+0}_{-13}$  &   43.77$^{+0.23}_{-0.52}$  \\
 A8  &     0.58$^{+0.09}_{-0.34}$   &     17.5$\pm$5.5    &     4.45$\pm$1.39  &  3.8$^{+0.9}_{-1.9}$  &    86$^{+20}_{-43}$ &   45.58$^{+0.20}_{-0.39}$  \\
 A9  &     1.00$^{+0.00}_{-0.33}$   &      6.7$\pm$4.1    &     1.15$\pm$0.71  &  4.7$^{+0.0}_{-2.0}$  &    30$^{+0}_{-19}$  &   43.60$^{+0.28}_{-0.96}$  \\
\enddata
\tablenotetext{a}{Broad-band (0.3--8 keV) X-ray counts.}
\tablenotetext{b}{Observed broad-band (0.3--8 keV) X-ray flux in 10$^{-15}$\ergcm2s.}
\tablenotetext{c}{\nh\ in the observer rest-frame in 10$^{22}$ cm$^{-2}$.}
\tablenotetext{c}{\nh\ in the source rest-frame in 10$^{22}$ cm$^{-2}$.}
\tablenotetext{d}{Absorption-corrected broad-band (0.3-8 keV) X-ray luminosity in \ergs, assuming H$_0$ = 71 \kmsMpc, $\Omega_{M}$=0.27 and $\Omega_{\Lambda}$=0.73.}
\end{deluxetable}

\begin{deluxetable}{lccccc}
\tabletypesize{\footnotesize}
\tablecaption{PAH Strengths and Luminosities of the Starburst Sample\label{tab4}}
\tablewidth{0pt}
\tablehead{
\colhead{ID}&
\colhead{flux density\tablenotemark{a}} &
\colhead{flux\tablenotemark{b}}&
\colhead{D\tablenotemark{c}} &
\colhead{L(PAH)\tablenotemark{d}}&
\colhead{L(cont)\tablenotemark{e}}
}
\startdata
 B1 & 1.7 & 2.3 & 14900  & 5.8 & 1.1\\
 B2 & 1.9 & 2.6 & 13900  & 5.7 & 1.0\\  
 B3 & 1.7 & 2.4 & 14000  & 5.2 & 0.9\\
 B4 & 2.5 & 3.7 & 12600  & 6.6 & 1.2\\
 B5 & 1.6 & 2.2 & 14600  & 5.3 & 1.0\\
 B6 & 1.4 & 2.8 &  6480  & 1.3 & 0.2\\
 B7 & 1.5 & 2.0 & 14600  & 4.9 & 0.9\\
 B8 & 2.0 & 2.7 & 14700  & 6.6 & 1.2\\
 B9 & 2.0 & 2.9 & 14200  & 6.5 & 1.2\\
\enddata

\tablenotetext{a}{Observed flux density in mJy at peak of redshifted 7.7\,\um PAH emission feature.}
\tablenotetext{b}{Observed flux $\nu$f$_{\nu}$ in units of 10$^{-13}$erg cm$^{-2}$s$^{-1}$ at redshifted peak of 7.7\,\um PAH emission feature.}
\tablenotetext{c}{Luminosity distance in Mpc.}
\tablenotetext{d}{Luminosity $\nu$L$_{\nu}$ in units of 10$^{45}$erg s$^{-1}$ at peak of 7.7\,\um PAH emission feature.}
\tablenotetext{e}{Luminosity $\nu$L$_{\nu}$ in continuum at 7.7\,\um in units of 10$^{45}$erg s$^{-1}$; continuum luminosity is determined using average ratio of peak flux at 7.7\,\um to underlying continuum flux of 5.5 derived from the average spectrum in Figure 10.}
\end{deluxetable}

\newpage
\clearpage
%
%
\begin{figure}
\figurenum{1}
\includegraphics[scale=1.0]{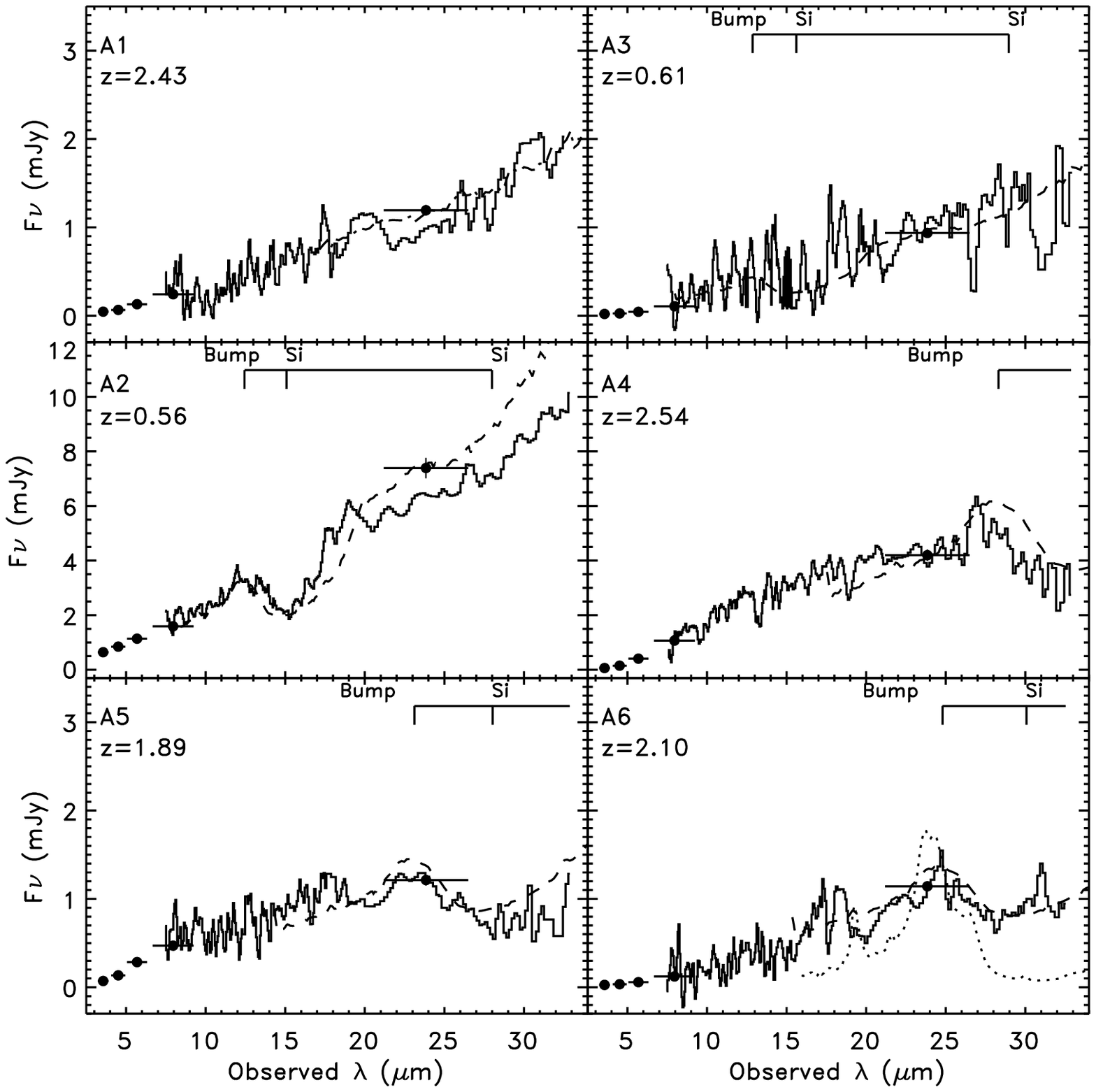}
\caption{Observed spectra of sources A1-A6 in Table 1, smoothed to approximate resolution of individual IRS orders (histogram), with Markarian 231 AGN template showing silicate absorption at the listed redshift (dashed curve) or a QSO power law for source A1 without indication of silicate absorption (dotted-dashed curve).  Source A6 also shows the NGC 3079 starburst template (dotted curve) as an illustration that possible emission features in this spectrum are not fit by the PAH template. Filled circles: fluxes and bandwidths of MIPS and IRAC measurements. The label "Si" shows the location of silicate absorption; the label "Bump" illustrates the location of the continuum peak at $\sim$ 8\,\um which arises in a heavily absorbed source.}

\end{figure}

\newpage
\clearpage
%
%
\begin{figure}
\figurenum{2}
\includegraphics[scale=1.0]{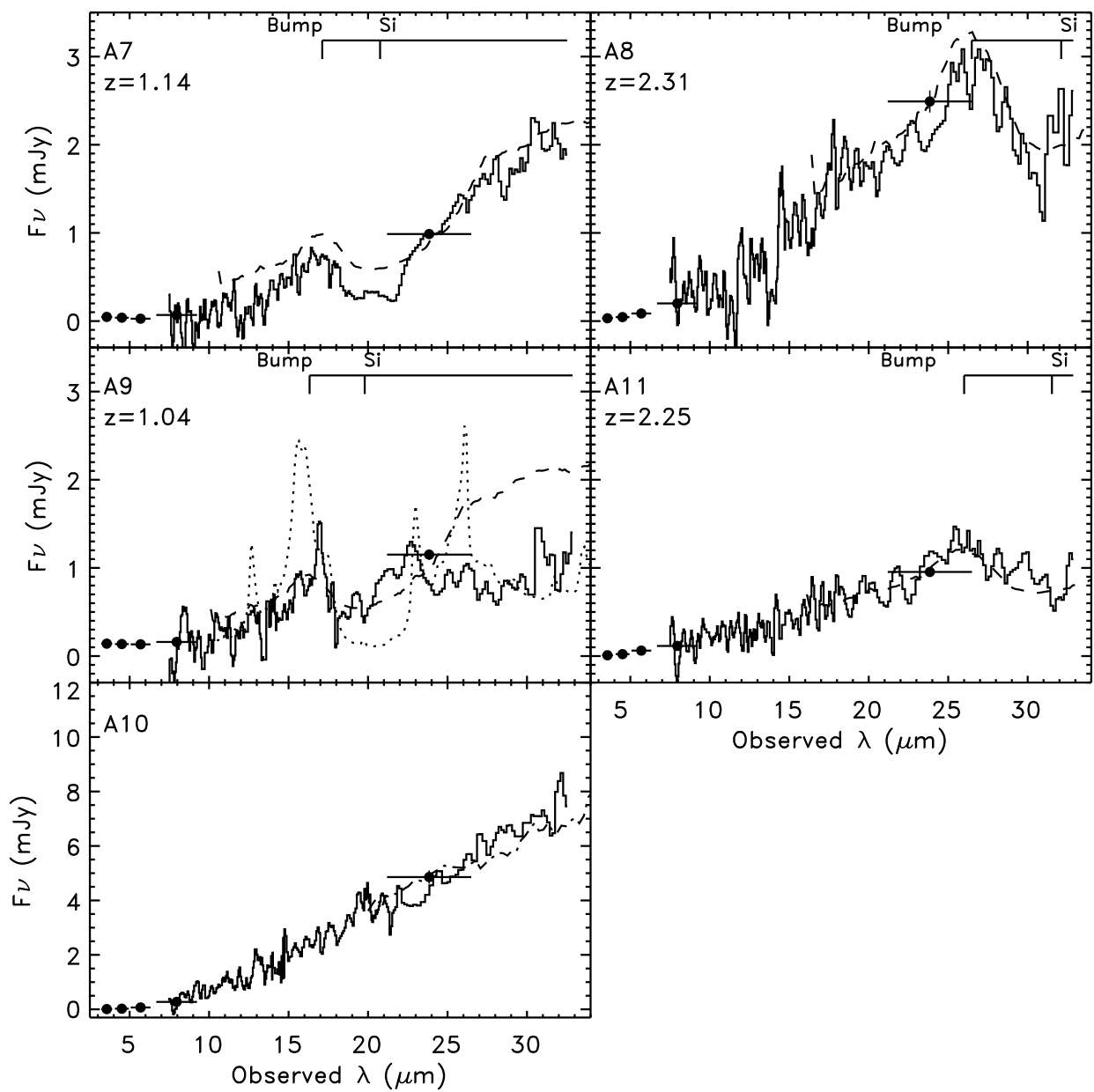}
\caption{Observed spectra of sources A7-A11 in Table 1, smoothed to approximate resolution of individual IRS orders (histogram), with Markarian 231 AGN template showing silicate absorption at the listed redshift (dashed curve) or a QSO power law for source A10 without absorption (dotted-dashed curve).  Source A9 also shows the NGC 3079 starburst template (dotted curve). Filled circles: fluxes and bandwidths of MIPS and IRAC measurements. The label "Si" shows the location of silicate absorption; the label "Bump" illustrates the location of the continuum peak at $\sim$ 8\,\um which arises in a heavily absorbed source.}

\end{figure}

\newpage
\clearpage

\begin{figure}
  \figurenum{3}
\includegraphics[scale=1.0]{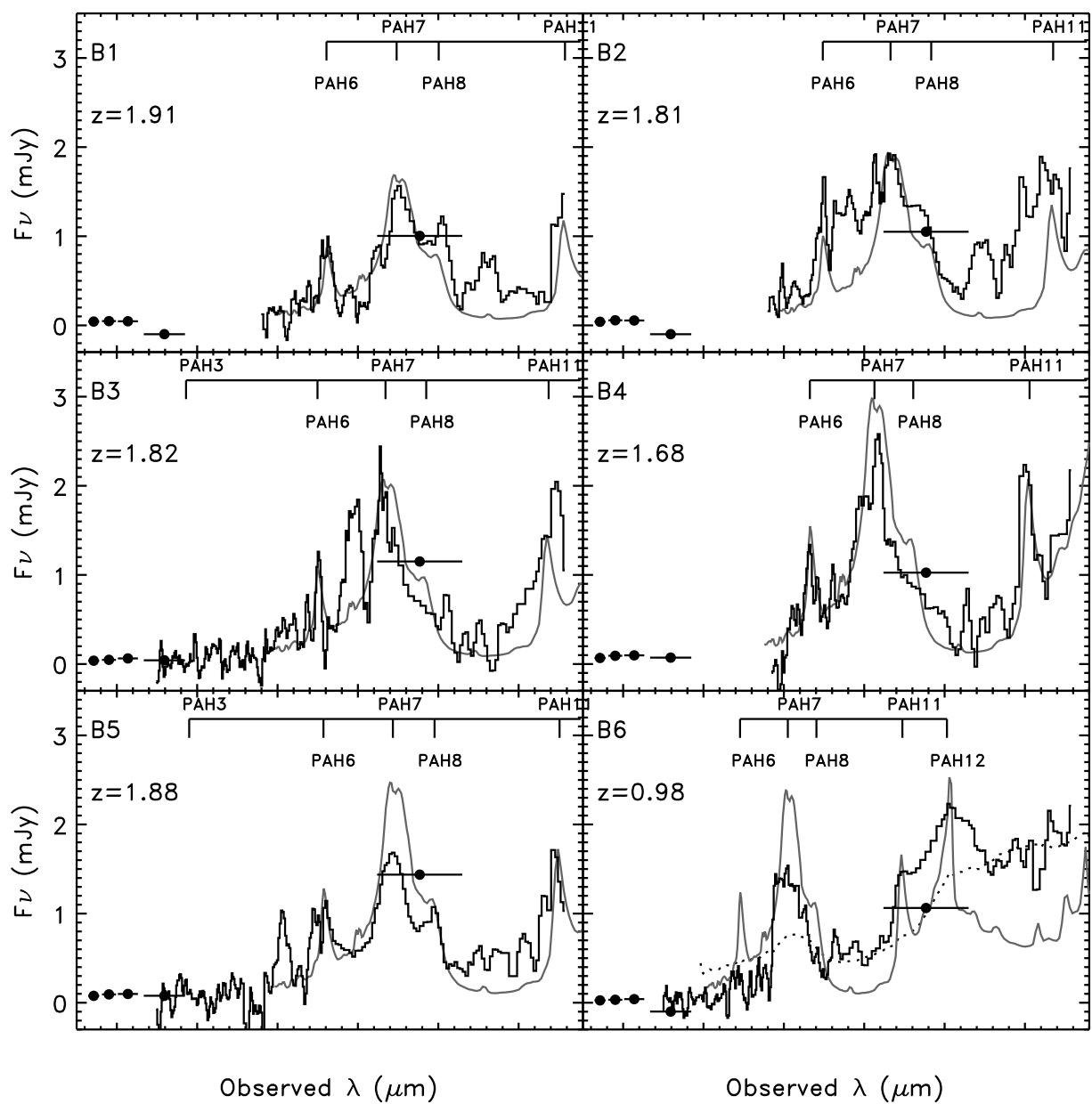}
\caption{Observed spectra of sources B1-B6 in Table 2, smoothed to approximate resolution of individual IRS orders (histogram), with NGC 3079 starburst template at the redshift based on PAH emission features (solid curve). Filled circles: fluxes and bandwidths of MIPS and IRAC measurements. Source B6 also shows the Markarian 231 template (dashed curve).}
\end{figure}

\newpage
\clearpage
%
%
\begin{figure}
\figurenum{4}
\includegraphics[scale=1.0]{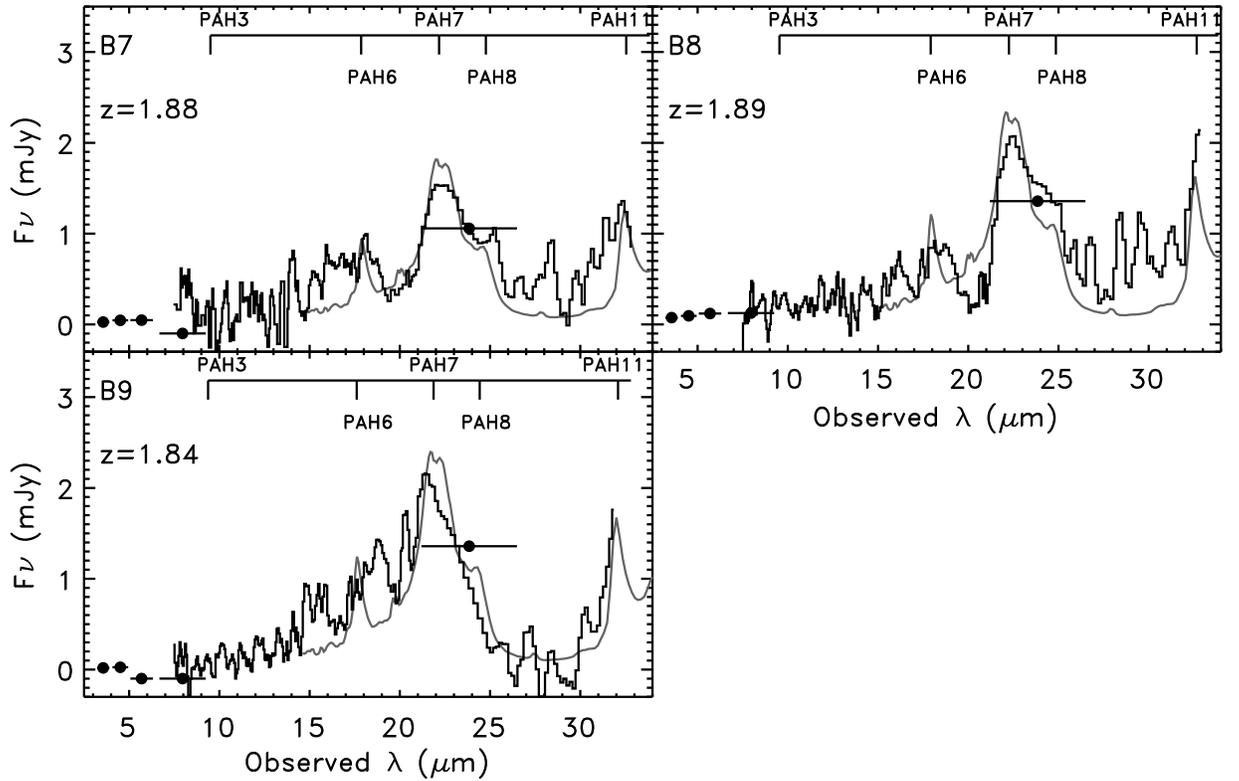}
\caption{Observed spectra of sources B7-B9 in Table 2, smoothed to approximate resolution of individual IRS orders (histogram), with NGC 3079 starburst template at the redshift based on PAH emission features (solid curve). Filled circles: fluxes and bandwidths of MIPS and IRAC measurements.}
  
\end{figure}

\newpage
\clearpage

\begin{figure}
  \figurenum{5}
\includegraphics[scale=1.0]{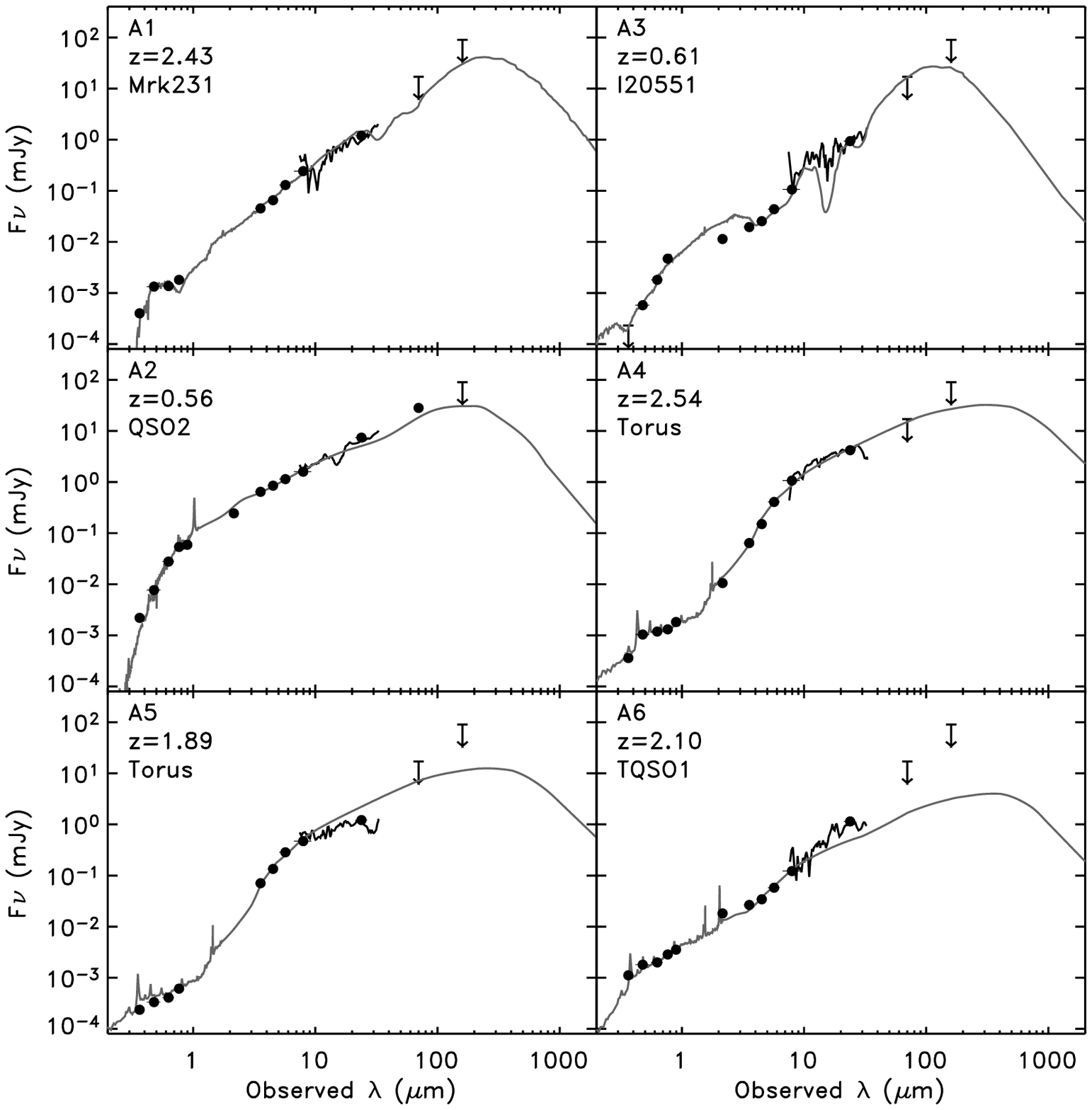}
\caption{Overall SEDs of AGN sources A1-A6 with SED template fits shown at the measured redshifts (solid curves).  Jagged curves: IRS spectra from Figure 1; filled circles: photometric points from Table 1; arrows are approximate 3 $\sigma$ upper limits (17 mJy at 70\,\um and 90 mJy at 160\,\um); all values and limits for photometry are from the SWIRE survey.}

\end{figure}

\newpage
\clearpage

\begin{figure}
  \figurenum{6}
\includegraphics[scale=1.0]{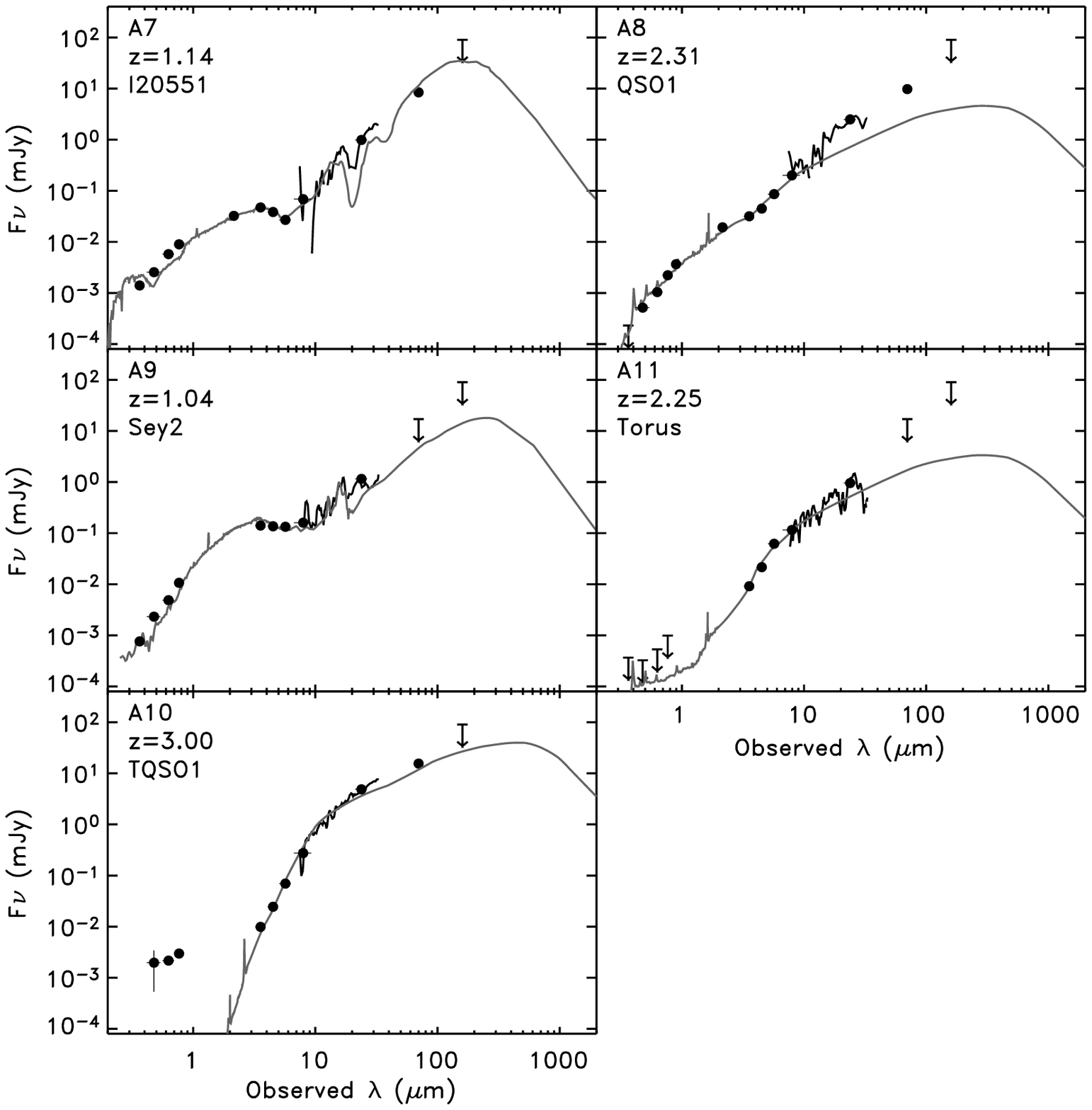}
\caption{Overall SEDs of AGN sources A7-A10 with SED template fits shown at the measured redshifts (solid curves).  Jagged curves: IRS spectra from Figure 2; filled circles: photometric points from Table 1; arrows are approximate 3 $\sigma$ upper limits (17 mJy at 70\,\um and 90 mJy at 160\,\um); all values and limits for photometry are from the SWIRE survey. Source A10 has no redshift estimate; the SED is shown at z = 3 for illustration.}

\end{figure}

\newpage
\clearpage


\begin{figure}
  \figurenum{7}
\includegraphics[scale=1.0]{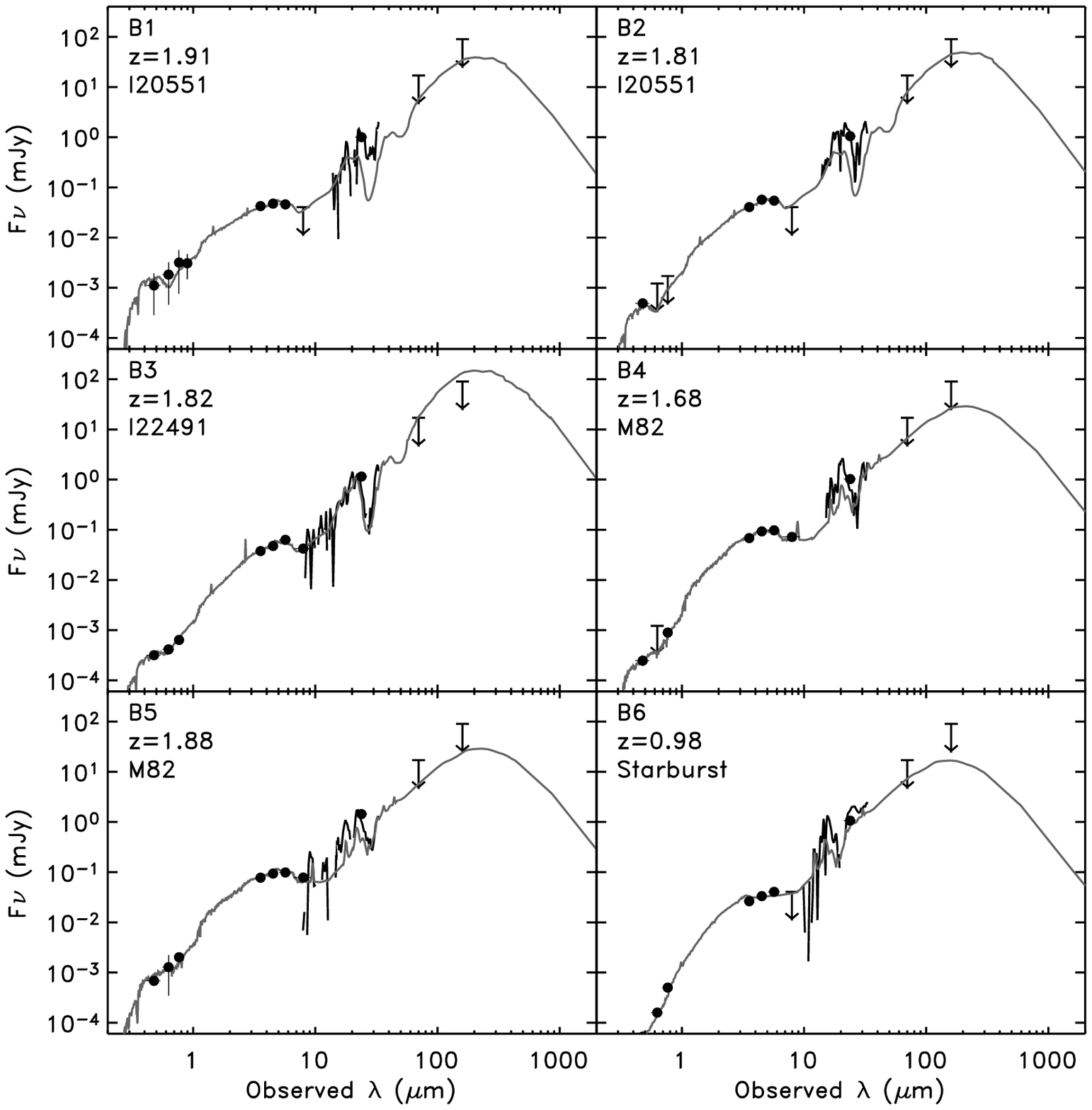}
\caption{Overall SEDs of starburst sources B1-B6 with SED template fits shown at the measured redshifts (solid curves).  Jagged curves: IRS spectra from Figure 3; filled circles: photometric points from Table 2; arrows are approximate 3 $\sigma$ upper limits (17 mJy at 70\,\um and 90 mJy at 160\,\um); all values and limits for photometry are from the SWIRE survey. The "hump" centered at observed wavelength $\sim$ 5\,\um is the signature of the stellar photospheric maximum at rest-frame $\sim$ 1.6\,\um.}
\end{figure}

\newpage
\clearpage

\begin{figure}
  \figurenum{8}
\includegraphics[scale=1.0]{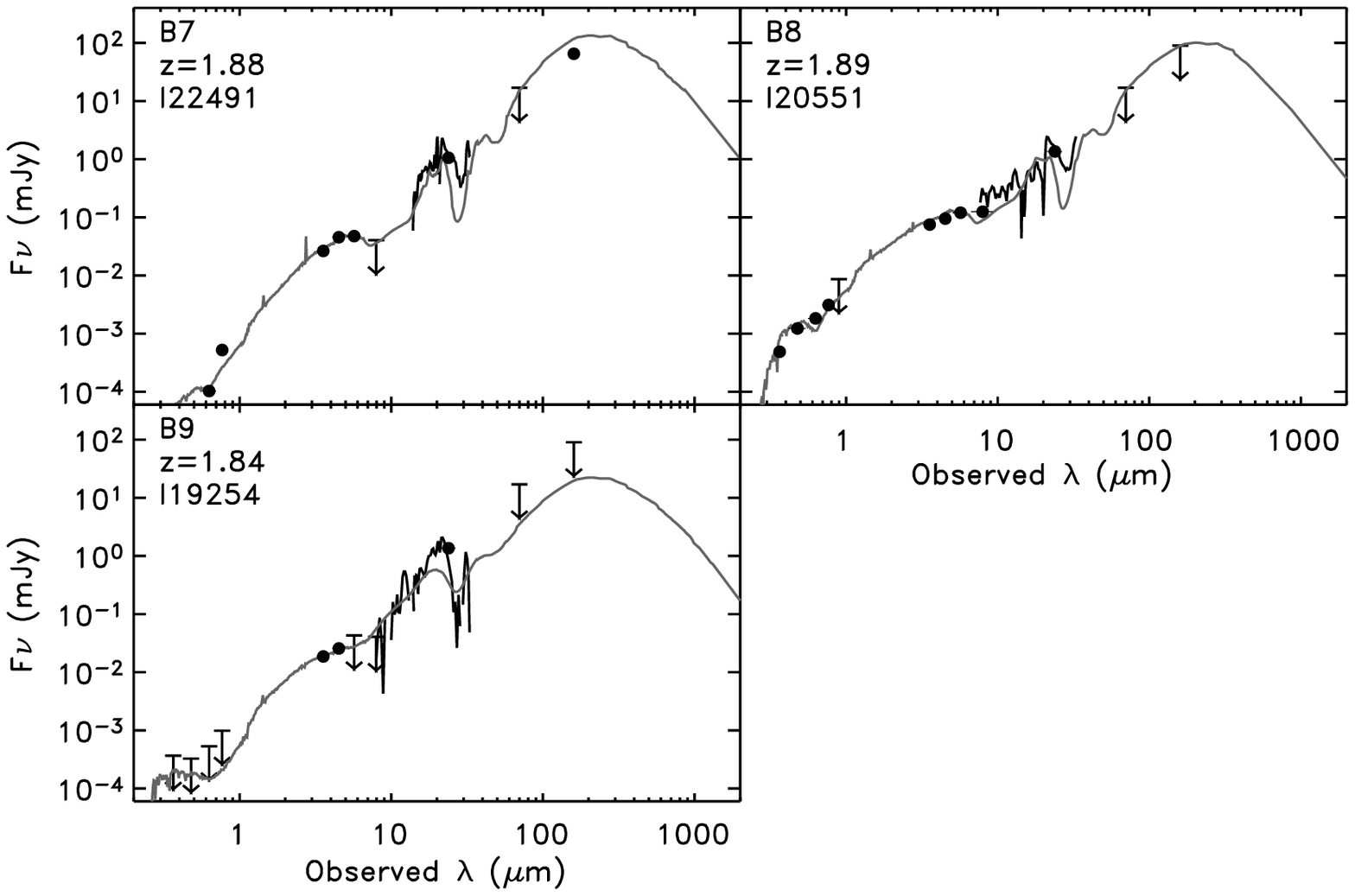}
\caption{Overall SEDs of starburst sources B7-B9 with SED template fits shown at the measured redshifts (solid curves).  Jagged curves: IRS spectra from Figure 4; filled circles: photometric points from Table 2; arrows are approximate 3 $\sigma$ upper limits (17 mJy at 70\,\um and 90 mJy at 160\,\um); all values and limits for photometry are from the SWIRE survey. The "hump" centered at observed wavelength $\sim$ 5\,\um is the signature of the stellar photospheric maximum at rest-frame $\sim$ 1.6\,\um.}

\end{figure}

\newpage
\clearpage

\begin{figure}
  \figurenum{9}
\includegraphics[scale=1.0]{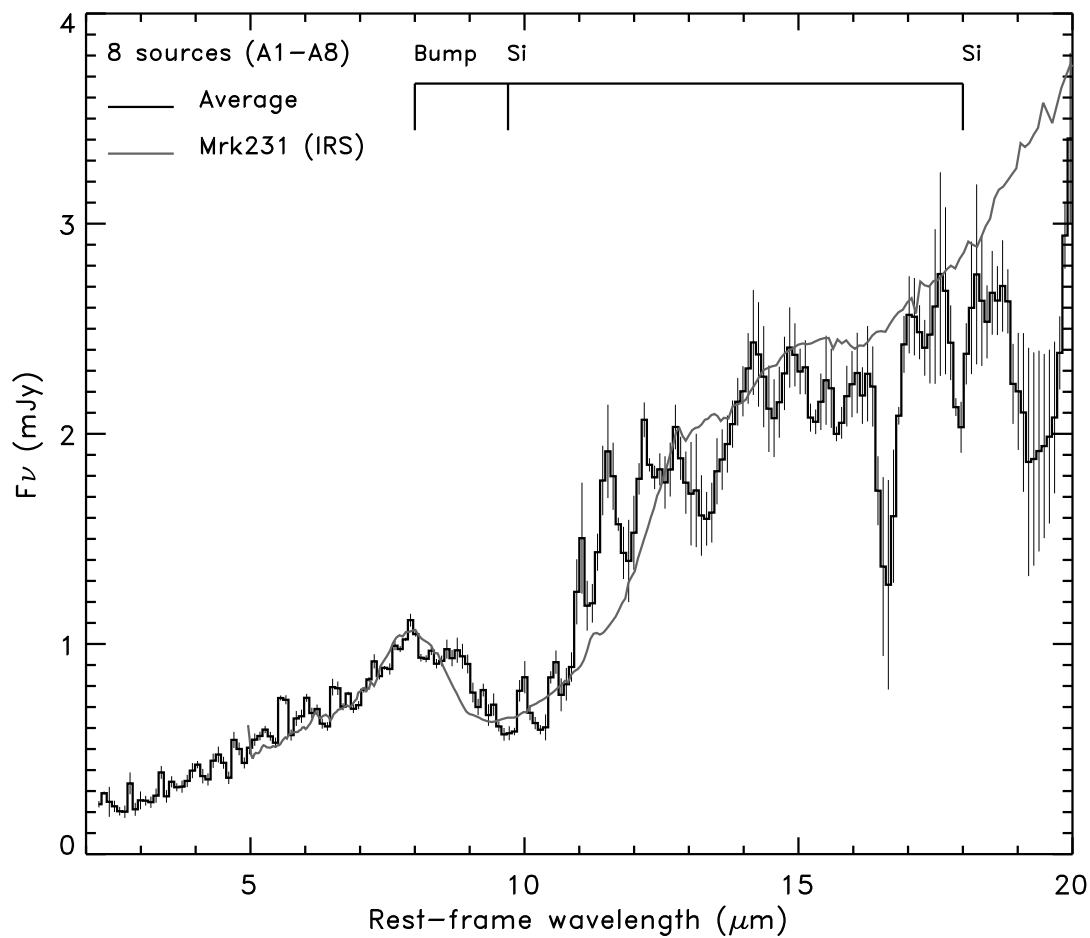}
\caption{Histogram: average rest-frame spectra of all AGN sources with redshifts in Table 1 normalized for flux density in interval 7.7\,\um to 8.0\,\um. Solid curve: template of Markarian 231; vertical bars: one sigma uncertainties for the average spectrum.}
\end{figure}

\newpage
\clearpage

\begin{figure}
  \figurenum{10}
\includegraphics[scale=1.0]{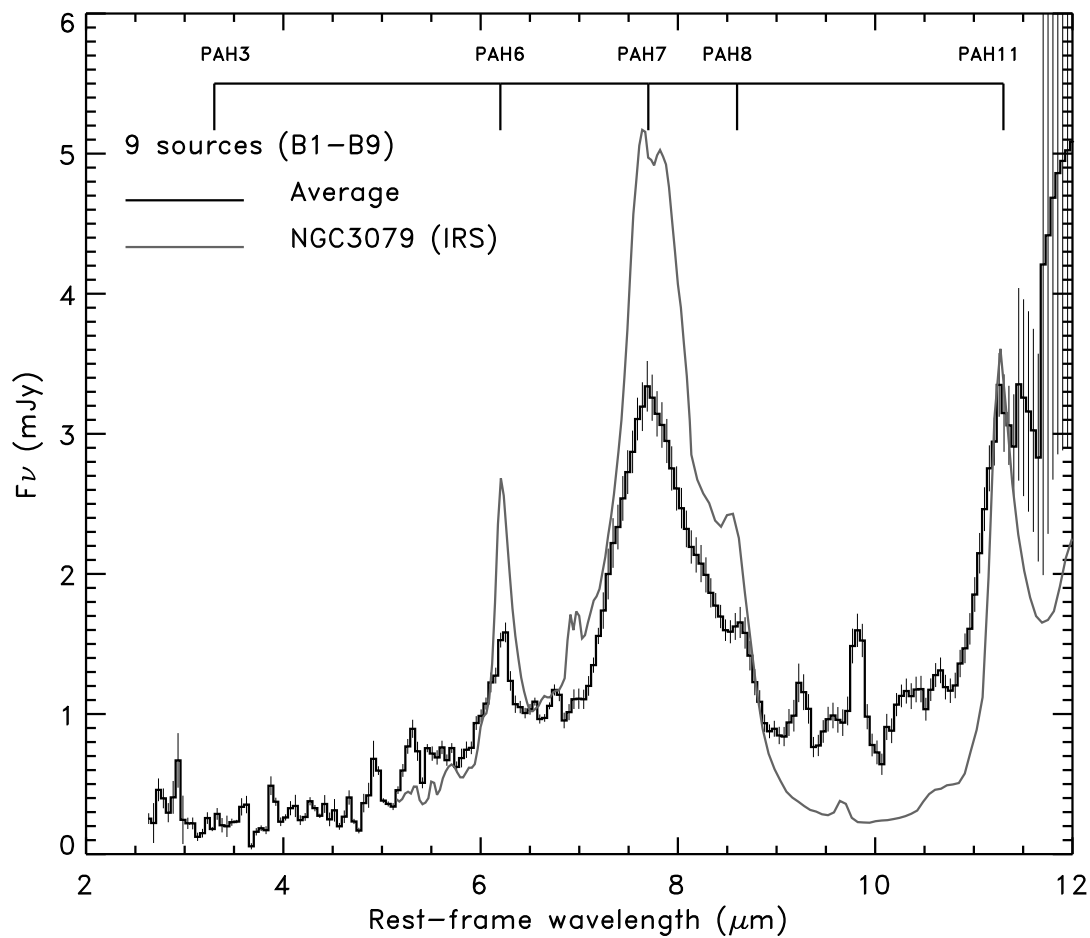}
\caption{Histogram: average rest-frame spectra of all starburst sources in Table 2 normalized for flux density in interval 6.3\,\um to 7.0\,\um. Solid curve: template of NGC 3079; vertical bars: one sigma uncertainties for the average spectrum.}
\end{figure}

\begin{figure}
  \figurenum{11}
\includegraphics[scale=1.0]{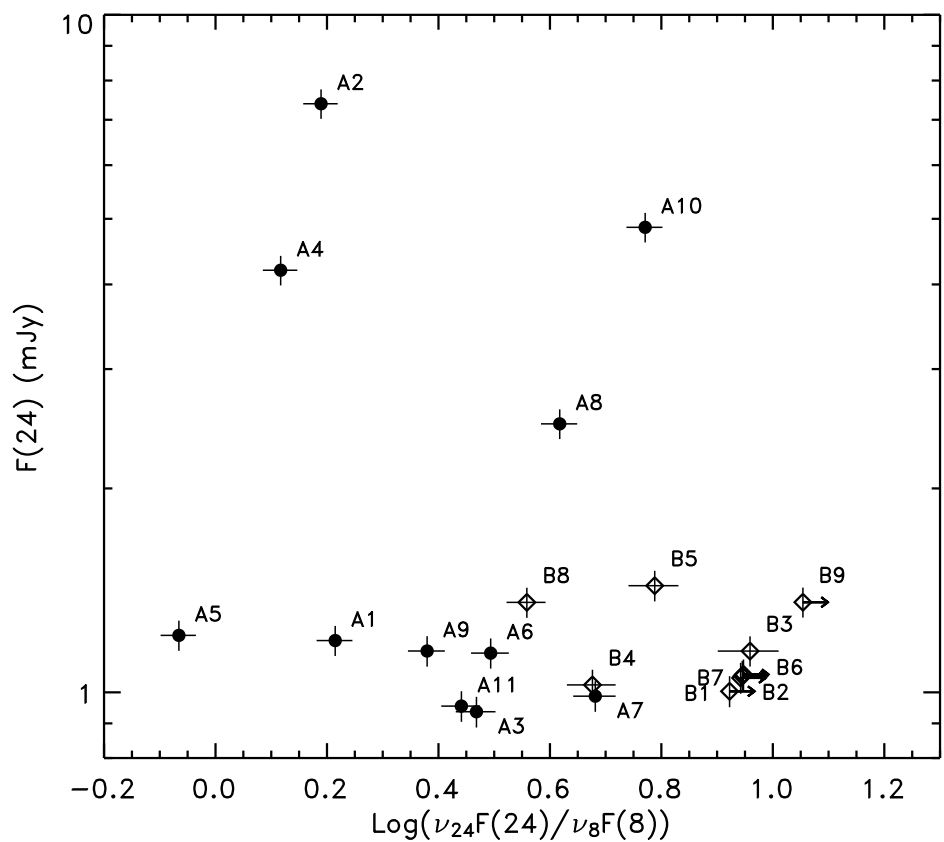}
\caption{Comparison of f$_{\nu}$ (24\,\um) to infrared spectral slope as measured by log[$\nu$f$_{\nu}$ (24\,\um)/$\nu$f$_{\nu}$ (8\,\um)]; filled circles: AGN in Table 1; open circles: starbursts in Table 2.} 
\end{figure}

\end{document}